\documentstyle[12pt]{article}

\newcommand{\be}{\begin{equation}}
\newcommand{\ee}{\end{equation}}
\newcommand{\bea}{\begin{eqnarray}}
\newcommand{\eea}{\end{eqnarray}}

\renewcommand{\theequation}{\arabic{section}.\arabic{equation}}

\begin{document}

\begin{titlepage}

\title{{\Large\bf 
Supersymmetric Quantization of \\
Anisotropic Scalar-Tensor Cosmologies}}
\author{{\large\sf James E. Lidsey}\thanks{E-mail: 
{\sf jel@maths.qmw.ac.uk}} 
\\ Astronomy Unit, School of Mathematical Sciences, \\
Queen Mary and Westfield, \\ 
Mile End Road, London E1 4NS, United Kingdom 
\and  
 {\large\sf  P. Vargas 
Moniz}\thanks{E-mail: {\sf pmoniz@mercury.ubi.pt}}~\thanks{URL: 
{\sf http://www.dfis.ubi.pt/$\sim$pmoniz}}~\thanks{Also at 
CENTRA, IST, Rua Rovisco Pais, 
1049 Lisboa Codex, Portugal}\\ 
Grupo de AsTrofisica e Cosmologia (GATC) \\
 Departamento de Fisica, 
Universidade da Beira Interior (UBI) \\
 Rua Marqu\^es d'Avila e Bolama, 
6200 Covilh\~a, Portugal }

\maketitle


\begin{abstract}
In this paper we 
show that the spatially homogeneous 
Bianchi type I and Kantowski--Sachs cosmologies 
derived from the Brans--Dicke theory of gravity 
admit a supersymmetric extension at the quantum level. 
Global symmetries 
in the effective one--dimensional actions characterize 
both classical and quantum solutions. 
A wide family 
of exact wavefunctions satisfying the 
supersymmetric constraints are found.
A connection with quantum wormholes is briefly 
discussed.

\end{abstract}

PACS number(s): 98.80.Hw, 04.60.Kz, 11.30.Pb

\end{titlepage}


\section{Introduction}

\setcounter{equation}{0}

\def\theequation{\thesection.\arabic{equation}}

\indent 

Quantum cosmology applies the fundamental 
principles of quantum physics to 
the entire universe. (For a 
review, see, e.g., Ref.  \cite{halliwell}). 
The wavefunction of the universe is a functional
on the configuration space (superspace) and 
obeys an infinite--dimensional partial differential 
equation -- the Wheeler--DeWitt equation
\cite{wheeler}. 
In view of the severe technical difficulties 
that arise in 
solving this equation, the normal procedure is to 
arbitrarily confine the fields 
to the neighbourhood 
of spatial homogeneity. Effectively, the infinite number of 
inhomogeneous modes and their interactions are truncated out
and the configuration space (minisuperspace) is 
therefore
finite--dimensional. 
The Wheeler--DeWitt equation then determines 
the evolution of the wavefunction on the 
minisuperspace
and 
a given trajectory mapped out by the wavefunction may 
be interpreted as a cosmological space--time. 

The validity of the minisuperspace approximation 
remains an open question to date. It is clearly inconsistent 
with the uncertainty principle since the amplitudes and 
momenta of the inhomogeneous modes are assumed to vanish 
simultaneously. 
Kucha\v{r} and Ryan   
addressed this question 
quantitatively within the context 
the Bianchi type IX cosmology  
and found that imposing additional symmetry on the model 
altered the nature of the physical predictions
\cite{kr} (see also Refs. \cite{kjap,kmg9}). 
On the 
other hand, 
Sinha and Hu  employed the 
techniques of coarse--graining and 
interacting field theory
to derive a condition that must be 
satisfied for the approximation 
to be justified \cite{hu}. 
Difficulties in coupling non--trivial spinor 
fields to highly symmetric spacetimes 
have also been  highlighted 
by Henneaux \cite{henn}.

Despite these uncertainties, however, 
the expectation is that the main  features of 
the wavefunction should be preserved in a more general 
analysis \cite{halliwellproc}. 
In principle, the wavefunction of the universe 
yields the probability that a 
spatial hypersurface evolves from a given initial state. 
However, ambiguities arise when attempting 
to invoke such an interpretation due to the 
hyperbolic nature of the Wheeler--DeWitt equation: a 
conserved current with a positive--definite probability 
density is not possible.

One possible resolution of this and related difficulties 
is to extend the standard quantization of the universe 
in a {\em supersymmetric} 
fashion. (For a review, see, e.g., Refs. 
\cite{paulo,eath}). Supersymmetry may help in the 
quantization of gravity for a number of reasons. 
Indeed, earlier work on supergravity 
theories \cite{A0} and recent developments in 
superstring theory \cite{GSW} and M--theory \cite{WittenM}
indicate that supersymmetry is a fundamental ingredient 
of any unified description of the fundamental 
interactions. Consequently, an analysis 
of the very early universe that includes supersymmetry 
is well motivated and 
a number of 
authors have developed models of the early universe 
where both quantum gravitational and supersymmetric effects are 
important \cite{paulo,eath,PDD,OO,onesusy,graham,Lidsey}. 
The advantage of such an approach is that the
quantum state of the universe, $\Psi$, is 
annihilated by 
supersymmetric constraints that 
are linear, first--order differential 
equations in the bosonic momenta 
variables. This is in contrast to 
nonsupersymmetric quantum cosmology, where 
a second--order Wheeler-DeWitt equation has to be solved
subject to suitable
boundary conditions \cite{hartlehawking,vilenkin,hp}. 
The supersymmetric algebra necessarily implies that 
$\Psi$ also obeys the Hamiltonian constraint and it is 
therefore sufficient to solve the first--order constraint 
equations
\cite{Teitelboim}. In many cases, 
this resolves ambiguities in the choice of 
factor ordering. Furthermore, supersymmetric quantum 
cosmology places the results of standard quantum 
cosmology in a wider perspective \cite{halliwell}
and a study of quantum 
minisuperspaces with supersymmetry may also provide (in spite 
of the obvious truncations) some helpful insights concerning 
the set of states that 
represent a complete formulation of quantum gravity 
with the other interactions. 

In recent years, two attractive (and possibly related)
approaches to 
supersymmetric quantum 
cosmology
have been developed. One approach is to begin with $N=1$ 
supergravity \cite{A0,PDD} in four dimensions and reduce the system 
to a one--dimensional model by invoking a  suitable 
homogeneous ansatz \cite{paulo,eath,OO,onesusy}. 
This leads to a minisuperspace 
with $N=4$ local supersymmetry. 
Alternatively, one may integrate a purely bosonic 
action over the spatial variables to derive a 
$(1+0)$--dimensional lagrangian and then 
perform a supersymmetric extension of the 
corresponding  Hamiltonian system by employing the 
quantization rules of the 
supersymmetric $\sigma$--model \cite{witten,ch,sigma,graham,Lidsey}. 
This results in an $N=2$ 
supersymmetry. In particular, this process could be related to the fact that 
any one-dimensional system is supersymmetric provided its 
ground state is normalizable \cite{REV1st}. Moreover, this 
technique can be 
generalized to higher 
dimensions by employing Darboux transformations
\cite{REV2nd}. 

In this paper we employ the latter approach 
to quantize spatially homogeneous 
cosmologies \cite{homogeneous} within the context of the 
Brans--Dicke theory of gravity \cite{bd}. 
The Brans--Dicke theory is relevant 
to the early universe and 
arises as the effective action of  
higher--dimensional gravity theories and, in particular, 
superstring theory \cite{GSW}. Moreover, 
the spatially flat and 
isotropic 
Brans--Dicke 
cosmology exhibits a discrete `scale factor 
duality' \cite{Lidsey,Veneziano91}. This symmetry forms 
the basis of the pre--big bang inflationary scenario 
\cite{Veneziano91,pbb} (for a review see, e.g., Ref. 
\cite{pbbreview})
and 
its origin can be traced to the T--duality of string theory
\cite{Giveon}. The consequences of  
scale factor duality for string quantum cosmology have been 
explored by a number of authors \cite{Lidsey,stringqc}. 
In particular,  supersymmetric 
quantum states have been found that respect the duality 
symmetry of the classical Hamiltonian \cite{Lidsey}. 
Our purpose in this paper is to perform  
a supersymmetric extension
of more general spatially homogeneous
cosmologies. 
Specifically, 
we consider the locally rotationally symmetric (LRS) Bianchi 
type I  model and the 
Kantowski--Sachs universe \cite{homogeneous}. Generalizations of scale 
factor duality have been 
shown to exist  in these models \cite{clancy,pietro} and 
we find supersymmetric wavefunctions that respect these 
symmetries. 
 
The paper is organized as follows. In 
Section II, the global symmetries of the actions and the supersymmetric 
quantization procedure are reviewed. 
The LRS Bianchi 
type I model is quantized in Section III and the vacuum 
Kantowski--Sachs model is quantized in Section IV. We conclude 
with a discussion in Section V, where we also comment on 
the possible 
relationship  between the supersymmetric minisuperspace 
extension \cite{graham,Lidsey,REV1st,REV2nd} 
and minisuperspaces retrieved from 
more general supergravity theories 
\cite{paulo,eath,onesusy}.

We assume throughout that $\hbar =1$.

\section{Supersymmetric Quantum Bianchi Cosmology}

\setcounter{equation}{0}

\def\theequation{\thesection.\arabic{equation}}

\subsection{Duality and the Wheeler--DeWitt Equation}

\indent

We consider the four--dimensional 
Brans-Dicke action given by 
\begin{equation}
S = \int d^4x \sqrt{-g} e^{-\Phi} 
\left[R - \omega (\nabla \Phi)^2 - 2\Lambda
\right] , 
\label{2p1eq}
\end{equation}
where $R$ is the Ricci curvature of the spacetime with 
metric $g_{\mu\nu}$, $g \equiv {\rm det}g_{\mu\nu}$ and 
$\Phi$ represents the Brans--Dicke (dilaton) field. The coupling 
between the scalar and tensor fields is parametrized by the 
constant, $\omega$, and $\Lambda$ is the 
cosmological constant in the gravitational sector of the theory. 
A consistent truncation of the string effective action is 
given by Eq. (\ref{2p1eq}) for $\omega =-1$ and $\Lambda < 0$
\cite{GSW}.
Dimensional reduction of higher--dimensional 
Einstein gravity on an isotropic, $d$--dimensional 
torus results in the above action, where $\omega =-1 +1/d$ and 
$\Phi$ determines the volume of the internal space \cite{Freund}.

The metric for the class of spatially homogeneous, 
LRS cosmological models with 
constant time hypersurfaces containing two--dimensional 
surfaces of constant curvature,  $k$, is 
given by \cite{homogeneous}
\begin{eqnarray}
ds^2 & =  & -N^2 dt^2 + e^{2\alpha - 4\beta} dr^2 
+
e^{2\alpha + 2\beta} d\Omega_{2,k}^2 
\label{2p2aeq} \\ 
 & =  & -N^2 dt^2 + a_1^2 dr^2 
+
a_2^2 d\Omega_{2,k}^2, 
\label{2p2beq}
\end{eqnarray}
where $N$ is the non--dynamical lapse function, $d\Omega_{2,k}^2$ 
is the unit metric on the constant curvature two--surfaces, 
$e^{3 \alpha (t)} \equiv a_1a_2^2$ determines the effective 
spatial volume of the universe and $\beta \equiv (1/3) [\ln (a_2 /
a_1)]$ determines the anisotropy of the model. The cases 
$k=\{ -1 , 0 , +1 \}$ correspond to the Bianchi type III, I 
and Kantowski--Sachs universes, respectively.
The geometry of the spatial sections of the Kantowsk--Sachs model 
is $S^1 \times S^2$. The symmetry group of these surfaces 
is of the Bianchi type IX, but only acts transitively 
on two--dimensional surfaces that foliate the three--space. 

Integrating over the spatial variables in Eq. (\ref{2p1eq}) 
for the metric ansatz (\ref{2p2aeq}) yields 
the minisuperspace action: 
\begin{equation}
\label{2p3eq}
S = \int dt 
N e^{3\alpha - \Phi}
\left[
-6 \frac{\dot{\alpha}^2}{N^2}
+
6\frac{\dot{\alpha}\dot{\Phi}}{N^2}
+
\omega\frac{\dot{\Phi}^2}{N^2}
+
\frac{\dot{\beta}^2}{N^2}
+
2k e^{-2\alpha - 2\beta} - 2 \Lambda
\right] .
\end{equation}

Introducing the new variables\footnote{We assume 
throughout this paper that $\omega > -4/3$.} 
\begin{eqnarray}
\sigma & = &  \sqrt{\frac{3+2\omega}{4+3\omega}}
(\Phi - 3\alpha)
\label{2p5eq} \\ 
u & = & 
\sqrt{\frac{8+6\omega}{2+\omega}} \left( 
\frac{1}{4+3\omega}\left[ \alpha +(1+\omega ) \Phi \right]
+ \beta \right) \\
v & = & 
\frac{1}{\sqrt{2+\omega}} \left[ 
\alpha +(1+\omega ) \Phi -2\beta \right]
\label{2p5eeq} 
\end{eqnarray}
implies that we may diagonalise the kinetic sector 
of 
the reduced action (\ref{2p3eq}): 
\begin{eqnarray}
S & = & \int dt
\left\{
\frac{1}{N}
e^{-\kappa\sigma}
\dot{u}^2
+
\frac{1}{N}
e^{-\kappa\sigma}
\dot{v}^2
-
\frac{1}{N}
e^{-\kappa\sigma}
\dot{\sigma}^2 \right. \nonumber \\
& + & \left. 2 N k 
e^{(C - \kappa) \sigma - G u} -  2 N \Lambda e^{-\kappa\sigma}
\right\},
\label{2p6eq}
\end{eqnarray}
where 
\begin{eqnarray}
\kappa \equiv \sqrt{\frac{4+3\omega}{3+2\omega}} \\
C \equiv  \frac{2(1+\omega)}{\sqrt{(3+2\omega)(4+3\omega)}} \\
G \equiv \sqrt{\frac{4+2\omega}{4+3\omega}}
.
\end{eqnarray}

Global symmetries in these models, corresponding 
to a generalization of scale 
factor duality, were uncovered in Ref. \cite{clancy}. 
The action (\ref{2p6eq}) is invariant 
under the discrete $Z_2$ `duality' symmetry 
\begin{equation}
\bar{u} = u, \quad \bar{v} = -v, \quad \bar{\sigma} =
\sigma 
\label{2p7eq}
\end{equation}
 and in terms of the original variables in Eq. (\ref{2p3eq}), 
this is equivalent to
\begin{eqnarray}
\bar{\alpha} & = & \frac{4+3\omega}{3(2+\omega)}\alpha
-
\frac{2(1+\omega)}{3(2+\omega)}\Phi
+
\frac{4}{3(2+\omega)}\beta
\label{2p8a1eq}
\\
\bar{\Phi} & = & -\frac{2}{2+\omega}\alpha
-
\frac{\omega}{2+\omega}\Phi
+
\frac{4}{2+\omega}\beta
\label{2p8a2eq}
\\
\bar{\beta} & = & \frac{2}{3(2+\omega)}\alpha
+
\frac{2(1+\omega)}{3(2+\omega)}\Phi
+
\frac{2+3\omega}{3(2+\omega)}\beta  .
\label{2p8a3eq}
\end{eqnarray}
The scale factors transform such that 
\begin{eqnarray}
\bar{a}_1 & = & a_1^{\frac{\omega}{2+\omega}} 
e^{-\frac{2(1+\omega)}{2+\omega}\Phi}
\label{2p8b1eq}
\\
\bar{a}_2 & = & a_2.
\label{2p8b2eq}
\end{eqnarray}
Thus, the scale factor $a_2$ is invariant under 
the symmetry transformation, whereas $a_1$ undergoes a direct 
inversion for the string inspired case, $\omega =-1$. 

The spatially flat Bianchi type I model $(k=0)$ also 
exhibits a global ${\rm SO}(2)$ symmetry that 
acts non--trivially on the variables $\{ u, v\}$:
\begin{eqnarray}
\bar{u} = \cos \theta u -\sin \theta  v \nonumber \\
\label{globalsym}
\bar{v} = \sin \theta u + \cos \theta v 
,
\end{eqnarray}
where $\theta$ is a constant. 
The equivalent transformations on the scale factors 
and dilaton field were presented in Ref. \cite{clancy}.
The variable, $\sigma$, transforms as a singlet under Eq. 
(\ref{globalsym}). 

The field equations for these models can be 
expressed in the form of an unconstrained Hamiltonian 
system, where the Hamiltonian vanishes.
The momenta conjugate to the variables $\{ u, v, \sigma \}$ 
are given by 
\begin{eqnarray}
\label{2p9aeq}
\pi_{u} &  = &  2 \dot{u} e^{-\kappa\sigma}
\nonumber \\
\pi_{v} & = &  2 \dot{v} e^{-\kappa\sigma}
\label{2p9beq}
\\
\pi_{\sigma} &  = &  - 2 \dot{\sigma} e^{-\kappa\sigma},
\label{2p9ceq}
\end{eqnarray}
from which the classical Hamiltonian constraint follows:
\begin{equation}
H = - \pi_{u}^2 
- 
\pi_{v}^2
+
\pi_{\sigma}^2
+
8k 
e^{(C - 2\kappa)\sigma}
e^{-G u}
-8\Lambda e^{-2\kappa\sigma}
 .
\label{2p10eq}
\end{equation}
Eq. (\ref{2p10eq}) may be written in the  
more compact form 
\begin{eqnarray}
H & = & G^{ab} \pi_a \pi_b + W(q^a),
\label{2p11aeq} \\
W(q^a) & = & - 8k 
e^{(C - 2\kappa)\sigma}
e^{-Gu}
+ 8\Lambda e^{-2\kappa\sigma} ,
\label{2p11beq}
\end{eqnarray}
where $q^a=(\sigma , u, v)$ $(a = 0, 1, 2)$
and 
$G^{ab} ={\rm diag} (-1 , 1 , 1)$ 
is the minisuperspace metric. 
By identifying the conjugate momenta with the operators 
$\pi_{q^a}=\pi_a = 
-i\partial /\partial q^a$ 
and neglecting ambiguities that arise in the 
factor ordering, we arrive at the
Wheeler-DeWitt equation:
\begin{equation}
\left[
-
\frac{\partial^2}{\partial \sigma^2}
+
\frac{\partial^2}{\partial u^2}
+
\frac{\partial^2}{\partial v^2}
+
 8k 
e^{(C - 2\kappa)\sigma}
e^{-G u}
- 8\Lambda e^{-2\kappa\sigma}
\right]
\Psi = 0.
\label{2p12eq}
\end{equation}

\subsection{Supersymmetric Quantum Cosmology}

\indent 

In this subsection we summarize the 
procedure for attaining a supersymmetric extension of the 
models. 
In general, such an 
extension of the system is possible if a 
solution, $I=I(q^a)$, can be found to the Euclidean Hamilton--Jacobi 
equation  \cite{witten,ch,graham}: 
\begin{equation}
\label{natural}
G^{ab}\frac{\partial I}{\partial q^a}
\frac{\partial I}{\partial q^b} = W (q^a)
\end{equation}
A quantum Hamiltonian, $\hat{H}$, 
may be defined by the conditions
\begin{equation}
\label{susyalg1}
2\hat{H} =[ Q, \bar{Q} ]_+ , \qquad Q^2 = \bar{Q}^2 =0 
\end{equation}
and 
\begin{equation}
\label{susyalg2}
[\hat{H} , Q ]_- = [\hat{H} , \bar{Q} ]_{-} =0
,
\end{equation}
where $Q$ is a non--Hermitian supercharge and $\bar{Q}$ 
is its adjoint. The functional forms of these supercharges 
are
\begin{equation}
\label{superform1}
Q =  \psi^a
\left( 
\pi_a + i\frac{\partial I}{\partial q^a}
\right)
\end{equation}
and 
\begin{equation}
\label{superform2}
\bar{Q} =  \bar{\psi}^a
\left( 
\pi_a - i\frac{\partial I}{\partial q^a}
\right)
, 
\end{equation}
respectively, where 
the corresponding fermionic (Grassmannian) variables are
defined by 
\begin{eqnarray}
\label{2p19eq}
\bar{\psi}^a = \theta^a, \qquad \psi^b = G^{ab}
\frac{\partial}{\partial \theta^a}  \\
\psi^a \psi^b + \psi^b \psi^a = 0 , \qquad 
\bar{\psi}^a \bar{\psi}^b + \bar{\psi}^b \bar{\psi}^a 
= 0 .
\end{eqnarray}

Eqs. (\ref{susyalg1}) and 
(\ref{susyalg2}) represent the algebra for a $N=2$ supersymmetry. 
For the three--dimensional minisuperspace that we are 
considering, the supersymmetric wavefunction can 
be expanded in terms of the Grassmann variables $\theta^a$: 
\begin{equation}
\Psi = A_+ + B_a\theta^a 
+ \frac{1}{2} \epsilon_{abc} C^c\theta^a
\theta^b 
+
A_-\theta^0 \theta^1 \theta^2,
\label{2p20eq}
\end{equation}
where the bosonic variables $\{ A_+, B_a, C_c, A_- \} $ are 
functions of the minisuperspace 
variables, $\epsilon_{abc}$ is totally 
antisymmetric on all its indices and 
$\epsilon_{012} \equiv +1$, etc. 
The supersymmetric wavefunction is then annihilated by the 
supercharges: 
\begin{eqnarray}
\label{super1}
Q \Psi = 0 \\
\label{super2}
\bar{Q} \Psi =0
\end{eqnarray}
and automatically satisfies 
the Hamiltonian constraint due to Eq. (\ref{susyalg1}).

The Euclidean Hamilton--Jacobi equation for the LRS Bianchi models 
we are considering
is given by 
\begin{equation}
-\left(
\frac{\partial I}{\partial \sigma}
\right)^2
+
\left(
\frac{\partial I}{\partial u}
\right)^2
+
\left(
\frac{\partial I}{\partial v}
\right)^2
= 
- 
 8k 
e^{(C - 2\kappa)\sigma}
e^{-Gu}
+ 8\Lambda e^{-2\kappa\sigma}.
\label{2p13eq}
\end{equation}

Thus, the problem of quantizing these models 
in a supersymmetric fashion involves 
finding a solution to Eq. (\ref{2p13eq})  and then 
solving the simultaneous constraints (\ref{super1}) and 
(\ref{super2}) subject to the ansatz (\ref{2p20eq}). 
Furthermore, since the models exhibit a classical, global 
symmetry, it is natural to consider those 
solutions to Eq. (\ref{natural}) that respect this symmetry. 

In the following Section we employ this method to quantize the 
LRS Bianchi I cosmology.

\section{Supersymmetric LRS Bianchi I Quantum Cosmology}

\indent 

\setcounter{equation}{0}

\def\theequation{\thesection.\arabic{equation}}

The Euclidean Hamilton--Jacobi equation for the LRS 
Bianchi type I cosmology is  
\begin{equation}
-\left(
\frac{\partial I}{\partial \sigma}
\right)^2
+
\left(
\frac{\partial I}{\partial u}
\right)^2
+
\left(
\frac{\partial I}{\partial v}
\right)^2
= 
 8\Lambda e^{-2\kappa\sigma}.
\label{2p14eq}
\end{equation}
A solution to Eq. (\ref{2p14eq}) that respects 
the global symmetry (\ref{globalsym}) and 
discrete ${\rm Z}_2$ symmetry (\ref{2p7eq})  of the reduced action 
(\ref{2p6eq}) 
is given by 
\begin{eqnarray}
\label{euclidean}
\Lambda < 0 & : & 
I = \mp \frac{1}{\kappa} \left[  \sqrt{A^2 -8\Lambda x^2} -A
{\rm cotanh}^{-1} \left( \frac{\sqrt{A^2 -8\Lambda x^2}}{A} \right) 
\right] + A \sqrt{u^2 +v^2} \label{euclidian1} \\
\Lambda > 0 & : & 
I = \mp \frac{1}{\kappa} \left[  \sqrt{A^2 -8\Lambda x^2} -A
{\rm tanh}^{-1} \left( \frac{\sqrt{A^2 -8\Lambda x^2}}{A} \right) 
\right] + A \sqrt{u^2 +v^2}, \label{euclidian2} 
\end{eqnarray}
where $A$ is an arbitrary constant and $x \equiv e^{-\kappa \sigma}$. 
In the $\Lambda > 0$ case one also requires
$\sigma \leq {\rm ln}(2\sqrt{2}\Lambda^{1/2}/A)$ if $A>0$ and 
$\sigma \geq {\rm ln}(-2\sqrt{2}\Lambda^{1/2}/A)$ for $A<0$.

Given the solution (\ref{euclidean}), 
we 
could in principle 
quantise the system in a manifestly supersymmetric fashion.
For simplicity, however,  we consider the 
case\footnote{This 
corresponds to the limit $\sigma \rightarrow -\infty$ and denotes 
a weak coupling regime for $\sigma$. In the strong limit 
($\sigma \rightarrow +\infty$), the term in $A$ dominates and 
$\Lambda$ can be positive. In the former situation, the `averaged' 
scale factor volume (represented by $\alpha$) and the 
dilaton are more important, while in 
the latter a large anisotropy in the spatial directions 
dominates. This  allows $\Lambda$ to be positive. In 
particular, one may have $\sigma \leftarrow +\infty$ and 
$u \rightarrow -\infty$, with, e.g., the 
singularity $a_2 \rightarrow 0$, $a_1 \rightarrow 
\infty$.}  
where $A=0$. 
The Euclidean action (\ref{euclidean}) then 
simplifies to 
\begin{equation}
\label{susysol}
I = \mp \frac{2\sqrt{-2 \Lambda}}{\kappa}
e^{-\kappa\sigma} 
\label{2p16eq}
\end{equation}
and it follows immediately from Eq. (\ref{2p16eq})
that we require $\Lambda \equiv -\lambda <0$ for consistency. 
The supercharges (\ref{superform1}) and 
(\ref{superform2}) are then given by 
\begin{equation}
Q  =  i\frac{\partial}{\partial \theta^0}
\frac{\partial}{\partial \sigma}
-
i\frac{\partial}{\partial \theta^1}
\frac{\partial}{\partial u}
-
i\frac{\partial}{\partial \theta^2}
\frac{\partial}{\partial v}
 \mp i 2\sqrt{2\lambda}e^{-\kappa\sigma}
\frac{\partial}{\partial \theta^0},
\label{2p17eq}
\end{equation}
and 
\begin{equation}
\bar{Q} =   -i\theta^0\frac{\partial}{\partial \sigma}
-
i\theta^1
\frac{\partial}{\partial u}
-
i\theta^2
\frac{\partial}{\partial v}
\mp i 2\sqrt{2\lambda}e^{-\kappa\sigma}
 \theta^0,
\label{2p18eq}
\end{equation}
respectively. 

The constraint (\ref{super2}) yields the  
set of coupled, first--order partial 
differential equations 
\begin{eqnarray}
-i\frac{\partial A_+}{\partial \sigma} 
& \mp & i 2\sqrt{2\lambda}
e^{-\kappa\sigma} A_+ = 0,
\label{2p21aeq} \\
-i\frac{\partial A_+}{\partial u}
& = & 0, 
\label{2p21beq} \\
-i\frac{\partial A_+}{\partial v}
& = & 0, 
\label{2p21ceq} \\
-i\frac{\partial B_1}{\partial \sigma} 
& + & 
i\frac{\partial B_0}{\partial u} 
\mp i 2\sqrt{2\lambda}
e^{-\kappa\sigma} B_1 = 0,
\label{2p21deq} \\
-i\frac{\partial B_2}{\partial \sigma} 
& + & 
i\frac{\partial B_0}{\partial v} 
\mp i 2\sqrt{2\lambda}
e^{-\kappa\sigma} B_2 = 0,
\label{2p21eeq} \\
-i\frac{\partial B_2}{\partial u} 
& + & 
i\frac{\partial B_1}{\partial v} 
= 0,
\label{2p21feq} \\
-i\frac{1}{2} 
\frac{\partial C^0}{\partial \sigma} 
& - & 
i\frac{1}{2}  
\frac{\partial C^1}{\partial u} 
- i\frac{1}{2}  
\frac{\partial C^2}{\partial v} 
\mp i  \sqrt{2\lambda}
e^{-\kappa\sigma} C^0 = 0 .
\label{2p21geq}
\end{eqnarray}
From the corresponding constraint (\ref{super1}), 
it follows that 
\begin{eqnarray}
i\frac{\partial A_-}{\partial \sigma} 
& \mp & i 2\sqrt{2\lambda}
e^{-\kappa\sigma} A_- = 0,
\label{2p22eeq} \\
i\frac{\partial A_+}{\partial u}
& = & 0, 
\label{2p22feq} \\
-i\frac{\partial A_-}{\partial v}
& = & 0, 
\label{2p22geq} \\
- \frac{i}{2}
\frac{\partial C^1}{\partial \sigma} 
& - & 
i\frac{\partial C^0}{\partial u} 
\pm i \sqrt{2\lambda}
e^{-\kappa\sigma} C^1 = 0,
\label{2p22deq} \\
i\frac{\partial C^2}{\partial \sigma} 
& + & 
i\frac{\partial C^0}{\partial v} 
\mp i 2\sqrt{2\lambda}
e^{-\kappa\sigma} C^2 = 0,
\label{2p22ceq} \\
i\frac{\partial C^2}{\partial u} 
& - & 
i\frac{\partial C^1}{\partial v} 
= 0,
\label{2p22beq} \\
i 
\frac{\partial B^0}{\partial \sigma} 
& -  & 
i 
\frac{\partial B^1}{\partial u} 
- i 
\frac{\partial B^2}{\partial v} 
\mp i 2 \sqrt{2\lambda}
e^{-\kappa\sigma} B^0 = 0.
\label{2p22aeq}
\end{eqnarray}

Eqs. (\ref{2p21aeq}), (\ref{2p21beq}) and  (\ref{2p21ceq}) 
immediately imply that 
\begin{equation}
A_+  = A_+^0 e^f
\label{2p23eq}
\end{equation}
with $A_+^0$ is an arbitrary constant and 
\begin{equation}
\label{fdefinition}
f \equiv 
\pm  \frac{2\sqrt{2\lambda}}{\kappa}
e^{-\kappa \sigma}.
\label{2p24eq}
\end{equation}
Similarly, we 
deduce from Eqs. 
(\ref{2p22eeq}), (\ref{2p22feq}) and (\ref{2p22geq})
that
\begin{equation}
A_-  = A_-^0 e^{-f} ,
\label{2p26eq}
\end{equation}
where $A_-^0$ is a second, arbitrary constant. 

To proceed in solving Eqs. 
(\ref{2p21deq}), (\ref{2p21eeq}), (\ref{2p21feq})
and (\ref{2p22aeq}), it proves convenient to 
redefine the 
functions $B_a$ as follows:
\begin{equation}
B_a \equiv \hat{B_a} e^{f}, \qquad a=(0,1,2)
.
\label{2p28eq}
\end{equation}
From the definition of $f(\sigma)$ given in Eq. 
(\ref{fdefinition}), 
it follows from Eqs. 
(\ref{2p21deq}), (\ref{2p21eeq}), (\ref{2p21feq}),
(\ref{2p22aeq}) and (\ref{2p28eq}) 
that 
\begin{eqnarray}
\frac{\partial \hat{B}_1}{\partial \sigma} 
& - & 
\frac{\partial \hat{B}_0}{\partial u} 
 = 0,
\label{2p30eq} \\
\frac{\partial \hat{B}_2}{\partial \sigma} 
& - & 
\frac{\partial \hat{B}_0}{\partial v} = 0,
\label{2p31eq} \\
\frac{\partial \hat{B}_2}{\partial u} 
& - & 
\frac{\partial \hat{B}_1}{\partial v} 
= 0,
\label{2p32eq} \\ 
\frac{\partial \hat{B}_0}{\partial \sigma} 
& -  &  
\frac{\partial \hat{B}_1}{\partial u} 
- 
\frac{\partial \hat{B}_2}{\partial v} 
- 2\kappa f  \hat{B}_0 = 0.
\label{2p33eq}
\end{eqnarray}
Similarly, by introducing the 
new set of variables $\hat{C}^b$ defined by 
\begin{equation}
C^b \equiv  \hat{C}^b e^{-f}
,
\label{2p34eq}
\end{equation}
we derive a new set of equations that are 
equivalent to Eqs. 
(\ref{2p21geq}), (\ref{2p22deq}), (\ref{2p22ceq})
and (\ref{2p22beq}): 
\begin{eqnarray}
\frac{\partial \hat{C}^1}{\partial \sigma} 
& + & 
\frac{\partial \hat{C}^0}{\partial u} 
 = 0,
\label{2p38eq} \\
\frac{\partial \hat{C}^2}{\partial \sigma} 
& + & 
\frac{\partial \hat{C}^0}{\partial v} = 0,
\label{2p37eq} \\
\frac{\partial \hat{C}^2}{\partial u} 
& - & 
\frac{\partial \hat{C}^1}{\partial v} 
= 0,
\label{2p36eq} \\ 
\frac{\partial \hat{C}^0}{\partial \sigma} 
& +  &  
\frac{\partial \hat{C}^1}{\partial u} 
+ 
\frac{\partial \hat{C}^2}{\partial v} 
+ 2\kappa f  \hat{C}^0 = 0.
\label{2p35eq}
\end{eqnarray}

By manipulating Eqs.  
(\ref{2p30eq})--(\ref{2p33eq})
we arrive at the following set of {\em decoupled}
equations\footnote{For example, Eq. 
(\ref{2p39eq}) is derived by applying the 
differential operator $\partial/\partial v$ on Eq. 
(\ref{2p33eq}), then acting 
on Eq.  (\ref{2p31eq}) 
with $\partial/\partial \sigma$ and on 
Eq. (\ref{2p32eq}) with 
$\partial/\partial u$. By employing 
Eq. (\ref{2p31eq}), we then arrive at Eq. 
(\ref{2p39eq}) above. A similar procedure leads to Eqs. 
(\ref{2p40eq}) and (\ref{2p41eq}).} 
\begin{eqnarray}
\frac{\partial^2 \hat{B}_2}{\partial \sigma^2} 
& - & 
2\kappa f 
\frac{\partial \hat{B}_2}{\partial \sigma} 
-
\frac{\partial^2 \hat{B}_2}{\partial u^2} 
-
\frac{\partial^2 \hat{B}_2}{\partial v^2} 
= 0,
\label{2p39eq} \\
\frac{\partial^2 \hat{B}_0}{\partial \sigma^2} 
& - & 
2\kappa f 
\frac{\partial \hat{B}_0}{\partial \sigma} 
+
2\kappa^2 f  \hat{B}_0
-
\frac{\partial^2 \hat{B}_0}{\partial u^2} 
-
\frac{\partial^2 \hat{B}_0}{\partial v^2} 
= 0, 
\label{2p40eq} \\
\frac{\partial^2 \hat{B}_1}{\partial \sigma^2} 
& - & 
2\kappa f 
\frac{\partial \hat{B}_1}{\partial \sigma} 
-
\frac{\partial^2 \hat{B}_1}{\partial u^2} 
-
\frac{\partial^2 \hat{B}_1}{\partial v^2} 
= 0.
\label{2p41eq}
\end{eqnarray}
Applying an equivalent technique to 
Eqs. (\ref{2p38eq})--(\ref{2p35eq}) results in a 
set of decoupled equations for the amplitudes 
$\hat{C}^c$: 
\begin{eqnarray}
-\frac{\partial^2 \hat{C}^2}{\partial \sigma^2} 
& - & 
2 \kappa f 
\frac{\partial \hat{C}^2}{\partial \sigma} 
+
\frac{\partial^2 \hat{C}^2}{\partial u^2} 
+ 
\frac{\partial^2 \hat{C}^2}{\partial v^2} 
= 0,
\label{2p44eq} \\
\frac{\partial^2 \hat{C}^0}{\partial \sigma^2} 
& + & 
2 \kappa f 
\frac{\partial \hat{C}^0}{\partial \sigma} 
-
2 \kappa^2 f  \hat{C}^0
-
\frac{\partial^2 \hat{C}^0}{\partial u^2} 
-
\frac{\partial^2 \hat{C}^0}{\partial v^2} 
= 0, 
\label{2p42eq} \\
-\frac{\partial^2 \hat{C}^1}{\partial \sigma^2} 
& - & 
2 \kappa f 
\frac{\partial \hat{C}^1}{\partial \sigma} 
+
\frac{\partial^2 \hat{C}^1}{\partial u^2} 
+
\frac{\partial^2 \hat{C}^1}{\partial v^2} 
= 0.
\label{2p43eq}
\end{eqnarray}

Eqs. (\ref{2p30eq})--(\ref{2p33eq}) can be solved 
analytically if $\hat{B}_{1,2}$ are independent 
of $\sigma$. Eqs. (\ref{2p39eq}) and 
(\ref{2p41eq}) then imply that these variables satisfy the 
two--dimensional Laplace equation, subject to the 
integrability condition (\ref{2p32eq}). 
Eqs. (\ref{2p30eq}) and 
(\ref{2p31eq}) further imply that $\hat{B}_0$ 
is independent of $\{ u ,v \}$ and 
consistency between Eqs. (\ref{2p33eq}) and 
(\ref{2p40eq}) results in a further 
integrability constraint
\begin{equation}
\label{further}
\frac{\partial \hat{B}_1}{\partial u} = -
\frac{\partial \hat{B}_2}{\partial v} 
.
\end{equation}
The functional form of $B_0$ follows 
immediately up on integration of 
Eq. (\ref{2p33eq}), $B_0 = e^{-f}$. 
It is interesting that this is also the wavefunction 
(\ref{2p26eq}) for the filled fermion sector. 
Similar conclusions follow for the functions $\hat{C}^c$.
If $\hat{C}^{1,2}$ are independent 
of $\sigma$, satisfy the two--dimensional Laplace equation, 
the integrability condition, $\partial \hat{C}^1 / \partial 
u = - \partial \hat{C}^2 / \partial 
v$, and Eq. (\ref{2p36eq}), then
the function $C^0$ is given by the wavefunction 
(\ref{2p23eq}) for the empty fermion sector, 
$C^0 = e^f$. 

Finally, it is interesting to 
compare the wavefunction (\ref{2p23eq}) for the empty 
fermion sector with the general solution to  the 
bosonic Wheeler--DeWitt 
equation (\ref{2p12eq}). When the wavefunction depends 
only on the variable $\sigma$, the general solution 
to Eq. (\ref{2p12eq}) is given by
\begin{equation}
\label{WDWsol}
\Psi = c_1 I_0 (f) +c_2 K_0 (f)
,
\end{equation}
where $I_0$ and $K_0$ are modified Bessel functions 
of the first and second kind with order zero, $f$ is defined in 
Eq. (\ref{2p24eq}) and $c_i$ are arbitrary constants. 
In the large argument limit, the modified Bessel 
function of the first kind asymptotes to the form 
$I_0 \propto f^{-1/2}\exp (f)$ and, consequently, there is a 
correlation, up to a 
negligible prefactor, with the fully bosonic component (\ref{2p23eq})
of the 
supersymmetric wavefunction. Indeed, the solution $A_+ = \exp (f)$ 
is an {\em exact} solution to the 
bosonic Wheeler--DeWitt equation if a suitable 
choice of factor ordering is made when 
identifying the momentum operator conjugate to the variable $\sigma$. 
In general, 
the ambiguity in the factor ordering can be accounted for 
\cite{hartlehawking} 
by identifying 
\begin{equation}
\pi_{\sigma}^2 = - e^{-p \sigma} \frac{\partial}{\partial  \sigma} 
e^{p \sigma}  \frac{\partial}{\partial \sigma} 
\end{equation}
for some constant $p$ in the classical Hamiltonian 
(\ref{2p10eq}). In this case, 
the corresponding 
Wheeler--DeWitt equation is then solved by 
Eq. (\ref{2p23eq}) for $p=\kappa$.

\section{Supersymmetric Kantowski--Sachs Quantum Cosmology}

\indent

In this Section, we consider the supersymmetric 
quantization of the vacuum Kantowski--Sachs, Brans--Dicke 
cosmology where $\Lambda =0$. The Wheeler--DeWitt and 
Euclidean Hamilton--Jacobi equations are given by 
\begin{equation}
\label{WDWKS}
\left[ -\frac{\partial^2}{\partial \sigma^2} + \frac{\partial^2}{\partial 
u^2} + \frac{\partial^2}{\partial v^2} +8e^{A\sigma +Bu} \right] 
\Psi =0
\end{equation}
and 
\begin{equation}
\label{EHJKS}
\left(
\frac{\partial I}{\partial \sigma}
\right)^2
-\left(
\frac{\partial I}{\partial u}
\right)^2 - \left( \frac{\partial I}{\partial v} \right)^2 
= 8
e^{A\sigma +B u}
,
\label{2p13neweq}
\end{equation}
respectively, where $A \equiv C -2\kappa$ and $B \equiv -G$. 

We now assume\footnote{We impose this restriction because it 
enables us to derive the 1-- and 2--fermion states 
analytically. Solutions that depend on $v$ 
can also be considered, although in such cases 
it is not possible to proceed
analytically.} the wavefunction does not depend on the 
variable $v$ and introduce `null' variables 
over the reduced $(1+1)$--dimensional 
minisuperspace: 
\begin{eqnarray}
\label{nullcoordinatepair}
s \equiv \frac{8}{A^2-B^2} \exp \left[ \frac{1}{2}
(A+B)(\sigma + u) \right] \nonumber \\
\tau \equiv \exp \left[ \frac{1}{2} (A-B)(\sigma - u)
\right]
.
\end{eqnarray}
The Wheeler--DeWitt equation (\ref{WDWKS}) transforms into 
the unit--mass Klein--Gordon equation 
\begin{equation}
\label{unitmass}
\left[ \frac{\partial^2}{\partial  s \partial \tau} 
-1 \right] \Psi =0
\end{equation}
and  particular solutions to Eq. (\ref{unitmass}) are
given by 
\begin{equation}
\label{basissol}
\Psi_{\mu} = e^{-i \mu s + i \tau /\mu}
,
\end{equation}
where $\mu$ is an arbitrary, complex constant. If ${\rm Im} \mu <0$, 
the modulus of the wavefunction is square--integrable.
The general 
solution to Eq. (\ref{unitmass}) may be expanded as a linear 
superposition of the family of solutions (\ref{basissol}): 
\begin{equation}
\label{genr}
\Psi_{\rm gen} =\int d^2 \mu F(\mu , \mu^* ) \Psi_{\mu}
\end{equation}
The function, $F$, represents a weighting function. If this is 
finite and only supported over a closed area of the ${\rm Im} \mu < 0$
sector of the complex $\mu$--plane, Cauchy's theorem implies that 
the integral in Eq. (\ref{genr}) may be reduced to a line integral over the 
real axis: 
\begin{equation}
\label{genKS1}
\Psi_{\rm gen} = \int^{+\infty}_{-\infty} 
d\mu M(\mu ) \Psi_{\mu}
,
\end{equation}
where 
$M(\mu )$ is an 
arbitrary function \cite{page}. 

The Euclidean Hamilton--Jacobi equation (\ref{EHJKS}) 
becomes
\begin{equation}
\label{EHJ}
\frac{\partial I}{\partial s}
\frac{\partial I}{\partial \tau} =1
\end{equation}
and admits the solutions 
\begin{equation}
\label{KSsolution}
I=  -bs -\frac{1}{b} \tau
.
\end{equation}
where $b= i\mu$. Eq. (\ref{KSsolution}) 
is invariant under the duality transformation 
(\ref{2p7eq}). Moreover, 
we see the exact solution (\ref{basissol})
to the Wheeler--DeWitt equation (\ref{WDWKS}) is also 
a WKB solution, $\Psi =\exp (\pm I)$, to the Euclidean 
Hamilton--Jacobi equation (\ref{EHJKS}).

In performing the supersymmetric 
quantization of this cosmology, it is convenient 
to diagonalise the minisuperspace metric. 
 The reason is that the 
Grassmannian variables should satisfy
the anticommuting relations  
$[\psi^a, \bar\psi^b]_+ = G^{ab}$. 
A non--diagonal minisuperspace metric would mean that 
fermionic states could not be clearly separated after 
the wavefunction has been annihilated by the supercharges. 
We therefore introduce the pair of variables
\begin{equation}
\label{TX}
T \equiv \frac{1}{2} (s +\tau ) , \qquad X \equiv \frac{1}{2}
( s -\tau )
\end{equation}
and this implies that the minisuperspace metric, defined in Eq. 
(\ref{2p11aeq}), has the non--trivial components
$G^{00}=-G^{11} = -(A^2-B^2)(T^2-X^2)/2$ and $G^{22} =1$.

The supercharge (\ref{superform1}) and 
its Hermitian conjugate (\ref{superform2}) are then 
given by
\begin{equation}
\label{superform1KS}
Q=-i G^{00} \frac{\partial}{\partial \theta^0} 
\frac{\partial}{\partial T} -i G^{11} 
\frac{\partial}{\partial \theta^1} \frac{\partial}{\partial X}
-i G^{22} \frac{\partial}{\partial \theta^2}
\frac{\partial}{\partial v} +iG^{00} 
\frac{\partial}{\partial \theta^0} \frac{\partial I}{\partial 
T} +iG^{11} \frac{\partial}{\partial \theta^1}
\frac{\partial I}{\partial X} 
\end{equation}
and 
\begin{equation}
\label{superform2KS}
\bar{Q} =-i \theta^0 \frac{\partial}{\partial T}
-i\theta^1 \frac{\partial}{\partial X} -i\theta^2 
\frac{\partial}{\partial v} -i\theta^0 
\frac{\partial I}{\partial T} -i \theta^1 
\frac{\partial I}{\partial X}
,
\end{equation}
respectively, where 
\begin{equation}
\label{inthefuture}
I =-\left( b+\frac{1}{b} \right) T- \left(
b- \frac{1}{b} \right) X  .
\end{equation}

For the supersymmetric wavefunction, we consider the ansatz
\begin{equation}
\label{waveKS}
\Psi = \alpha_+ +\beta_b \theta^b +\frac{1}{2} \epsilon_{abc}
\gamma^c \theta^a \theta^b + \alpha_- \theta^0 \theta^1 
\theta^2
,
\end{equation}
where $\{ \alpha_{\pm} , \beta_b , \gamma^c \}$ are bosonic 
functions of the minisuperspace variables. 
The annihilation of the wavefunction (\ref{waveKS}) 
by the supercharges (\ref{superform1KS}) 
and (\ref{superform2KS}) yields 
the set of coupled, first--order 
partial differential equations 
\begin{eqnarray}
\label{set1}
G^{00} \frac{\partial \beta_0}{\partial T} 
+ G^{11}\frac{\partial \beta_1}{\partial X} 
+G^{22} \frac{\partial \beta_2}{\partial v} +G^{00}
\left( b+\frac{1}{b} \right) \beta_0 -G^{11} \left( 
-b +\frac{1}{b} \right) \beta_1 =0 \\
\label{set1a}
-G^{11}  \frac{\partial \gamma^2}{\partial X} 
+ G^{22}  \frac{\partial \gamma^1}{\partial v}
+G^{11} \left( -b +\frac{1}{b} \right)  \gamma^2 =0 \\
\label{set1b}
G^{00}  \frac{\partial \gamma^2}{\partial T} 
-G^{22}  \frac{\partial \gamma^0}{\partial v}
+ G^{00} \left( b+\frac{1}{b} \right)  \gamma^2 =0 \\
\label{set1c}
- G^{00}  \frac{\partial \gamma^1}{\partial T} 
+G^{11}  \frac{\partial \gamma^0}{\partial X}
-G^{00} \left( b+\frac{1}{b} \right)  \gamma^1  - G^{11} 
\left( -b +\frac{1}{b} \right)  \gamma^0 =0 \\
\label{set1d}
\frac{\partial \alpha_-}{\partial v} =0 \\
\label{set1e}
\frac{\partial \alpha_-}{\partial T} + \left( 
b+\frac{1}{b} \right) \alpha_- =0 \\
\label{set1f}
\frac{\partial \alpha_-}{\partial X} 
+ \left( b -\frac{1}{b} \right) \alpha_- =0 \\
\label{set1g}
\frac{\partial \alpha_+}{\partial T} - \left( b +\frac{1}{b} 
\right) \alpha_+ =0 \\
\label{set1h}
\frac{\partial \alpha_+}{\partial X} - \left( b-\frac{1}{b} 
\right) \alpha_+ =0 \\
\label{set1i}
\frac{\partial \alpha_+}{\partial v} =0 \\
\label{set1j}
\frac{\partial \beta_1}{\partial T} -\frac{\partial \beta_0}{\partial X} 
- \left( b +\frac{1}{b} \right) \beta_1 + \left( b -\frac{1}{b}
\right) \beta_0 =0 \\
\label{set1k}
\frac{\partial  \beta_2}{\partial T} -
\frac{\partial \beta_0}{\partial v}  -\left( b+\frac{1}{b} 
\right) \beta_2  =0 \\
\label{set1m}
\frac{\partial \beta_2}{\partial X} -
\frac{\partial \beta_1}{\partial v} -\left( 
b -\frac{1}{b} \right) \beta_2 =0 \\
\label{set2}
 \frac{\partial \gamma^0}{\partial T} 
+  \frac{\partial \gamma^1}{\partial X} 
+  \frac{\partial \gamma^2}{\partial v}
-\left( b+\frac{1}{b} \right)  \gamma^0 
-\left( b -\frac{1}{b} \right)  \gamma^1 =0
.
\end{eqnarray}

The wavefunctions for the empty and filled fermion sectors are 
readily deduced: 
\begin{equation}
\label{emptyfilled}
\alpha_{\pm} = e^{ \mp I} 
,
\end{equation}
where $I$ is given by Eq. (\ref{inthefuture}). 
To solve the remaining equations, 
we
assume that the amplitudes $\{ \beta_b , \gamma^c \}$ are 
independent of the variable, $v$. In this case, 
Eqs. (\ref{set1k}) and (\ref{set1m}) yield the general 
solution, $\beta_2 = \exp (-I)$, modulo a 
constant of proportionality, where $I$ is given by 
the Euclidean action (\ref{KSsolution}). 
Likewise, 
Eqs. (\ref{set1a}) and (\ref{set1b}) imply 
that $\gamma^2 = \exp (I)$.

The wavefunctions for the one--fermion sector are completely 
determined by solving Eqs. (\ref{set1}) and (\ref{set1j}). 
To proceed, it is convenient to 
transform back to the null coordinate pair 
$(s , \tau )$ defined in Eq. (\ref{nullcoordinatepair}). 
In terms of these variables, we find that 
\begin{eqnarray}
\label{beta0}
\frac{\partial \left( \beta_0 -\beta_1 \right)}{\partial s}
+\frac{1}{b} \left( \beta_0 +\beta_1 \right) =0 \\
\label{beta1}
\frac{\partial \left( \beta_0 +\beta_1  \right)}{\partial \tau}
+b \left( \beta_0 -\beta_1 \right)
=0
.
\end{eqnarray}
Defining $Y \equiv \beta_0 -\beta_1$ and 
$Z \equiv b^{-1}(\beta_0 +\beta_1)$ implies that 
Eqs. (\ref{beta0}) and (\ref{beta1}) may be 
expressed in the more compact form: 
\begin{equation} 
\label{YZ}
\frac{\partial Y}{\partial s} =-Z , \qquad 
\frac{\partial Z}{\partial \tau} =-Y
 .
\end{equation}
Differentiating the first constraint in 
Eq. (\ref{YZ}) with respect to $\tau$ and 
substituting in the second condition implies 
that both $\beta_{0,1}$ satisfy the unit--mass 
Klein--Gordon equation, i.e., the bosonic 
Wheeler--DeWitt equation
(\ref{unitmass}). Thus, although 
these amplitudes satisfy 
the same equation as 
the wavefunction 
that arises in the standard 
quantum cosmological approach, the 
supersymmetry imposes strong constraints,
as summarized in Eq. 
(\ref{YZ}),  
on the functional form of the 
solutions that can arise. One class of allowed solution 
is given by $Y =\exp (-I)$ and $Z = -b Y$, 
where 
$I$ is given by Eq. (\ref{KSsolution}).

It now only remains to solve Eqs. 
(\ref{set1c}) and (\ref{set2}) in order 
to determine the two--fermion sector of the 
supersymmetric wavefunction. By combining 
and subtracting these two equations, 
we find that 
\begin{eqnarray}
\label{gammaadd}
\frac{\partial \left( \gamma^0 +\gamma^1 \right)}{\partial 
s} -\frac{1}{b} \left( \gamma^0 - \gamma^1 \right) =0 \\
\label{gammasubtract}
\frac{\partial \left( \gamma^0 -\gamma^1 \right)}{\partial 
\tau} -b  \left( \gamma^0 + \gamma^1 
\right) =0.
\end{eqnarray}
Defining $R \equiv \gamma^0 +\gamma^1$ and 
$W \equiv b^{-1}(\gamma^0 -\gamma^1)$ implies that 
Eqs. (\ref{gammaadd}) and (\ref{gammasubtract}) are 
equivalent to: 
\begin{equation} 
\label{RW}
\frac{\partial R}{\partial s} =W , \qquad 
\frac{\partial W}{\partial \tau} =R
.
\end{equation}
Thus, the amplitudes $\gamma^{0,1}$ also 
satisfy the unit--mass Klein--Gordon equation. We find that 
one class of solution consistent with Eq. (\ref{RW}) 
is given by $R=\exp (I)$ and $W =-bR$.
 To summarize, therefore, the 
supersymmetric wavefunction that we have found for the 
vacuum Brans--Dicke, Kantowski--Sachs cosmology 
is given by
\begin{eqnarray}
\label{susywavefunction}
\Psi = e^{-I} + \beta_0 \theta^0
+\beta_1 \theta^1 + e^{-I} \theta^2 \nonumber \\
+ 
\gamma^0 
 \theta^1 \theta^2 - \gamma^1  \theta^0 
\theta^2 
+ e^{I} \theta^0 \theta^1
+ e^I \theta^0 \theta^1 \theta^2
,
\end{eqnarray}
where $\beta_{0,1}$ and $\gamma^{0,1}$ 
satisfy the unit--mass, Klein--Gordon 
equation (\ref{unitmass}) 
subject to the integrability conditions (\ref{YZ})
and (\ref{RW}).

\section{Discussion}

\indent

In this paper, we have 
considered an $N=2$ supersymmetric quantization 
of the 
LRS Bianchi type I and 
Kantowski--Sachs, Brans-Dicke cosmologies. 
In the former case, 
we found that a supersymmetric quantization is possible if 
a negative cosmological constant is introduced into 
the gravitational sector of the theory. 
For the Kantowski--Sachs 
universe, the existence of such a term is not 
necessary, because this model has positive spatial 
curvature. In both models, supersymmetric 
quantum states were found for a given 
solution to the Euclidean Hamilton--Jacobi 
equation. Furthermore, these 
wavefunctions 
respect
a global scale factor 
duality symmetry of the respective classical 
Hamiltonians. 

Having found particular solutions to the supersymmetric  
quantum constraints, the immediate question that 
arises is the nature of the boundary conditions that 
such solutions satisfy. In general, the supersymmetric 
Hamiltonian has a spin term with a coefficient 
determined by the solution to the Euclidean 
Hamilton--Jacobi equation. This term implies that 
it is very difficult to complete a supersymmetric extension of the 
system with complex or imaginary solutions
\cite{graham}. Thus, 
the boundary conditions that are 
typically most relevant 
in this quantization scheme 
are those due to Hartle and Hawking \cite{hartlehawking} 
and to Hawking and Page \cite{hawking,hp}. 

In particular, 
it is natural to consider whether 
the Kantowski--Sachs wavefunction derived above 
satisfies the Hawking--Page boundary conditions 
relevant to a
wormhole configuration \cite{hawking,hp}. Classically, a wormhole 
represents an instanton solution 
of the Euclidean field equations \cite{classworm,hawking}. 
At the quantum 
level, such a state
may be interpreted as a solution to the 
Wheeler--DeWitt equation. The 
appropriate 
boundary conditions 
that must be satisfied are 
that the wavefunction should be regular, in the 
sense that it does 
not oscillate an infinite number of times, when the 
three--metric degenerates and that it 
should be exponentially damped when the three--geometry 
tends to infinity \cite{hawking,hp}. 

The anisotropic geometry, 
$S^1 \times S^2$, of the  Kantowski--Sachs model
implies that there are different types of possible wormholes
\cite{kl,cg}. 
These have been studied by Campbell and Garay within the 
context of Einstein gravity minimally coupled to a 
massless scalar field \cite{cg}.  
The geometry of the spacetime asymptotes to ${\rm R}^3 \times 
S^1$ if the radius of the circle, $\tilde{a}_1$, tends to a constant 
as the radius of the two--sphere diverges. 
Alternatively, if the volume of the two--sphere tends 
to a constant as $\tilde{a}_1 \rightarrow \infty$, the geometry is 
${\rm R}^2 \times S^2$. 
The wavefunction representing the ground state of each of these 
wormholes is the path integral over all metrics that 
asymptotically 
have these geometries and over all matter configurations that 
vanish at infinity. For the ${\rm R}^3 \times S^1$ wormhole, 
the wavefunction is given by $\Psi \propto \exp (-4
\tilde{a}_1 \tilde{a}_2)$ in the
asymptotic limit. The corresponding limit for the ${\rm 
R}^2 \times S^2$ wormhole is $\Psi \propto \exp ( -
\tilde{a}_1^2)$.

After transforming 
back to the original variables of Section II, 
we find that 
the bosonic component of the supersymmetric 
Kantowski--Sachs wavefunction 
(\ref{susywavefunction}) does not 
asymptote to either of these forms. Its interpretation 
as a quantum wormhole is therefore not clear. However, 
a further solution to the Euclidean 
Hamilton--Jacobi equation (\ref{EHJKS}) that respects
the scale factor duality (\ref{2p7eq}) of 
the classical action is given by
\begin{equation}
\label{wormholeaction}
I = \left( \frac{32}{A^2- B^2} \right)^{1/2} e^{(A \sigma 
+ B u )/2}
.
\end{equation}
Consequently, a supersymmetric quantization may be performed with 
this solution. Due to the non--trivial functional form 
of Eq. (\ref{wormholeaction}), however, it has not 
been possible to find 
analytical solutions for 
the intermediate fermionic sectors. On the other 
hand, the empty fermion 
sector is given by $\Psi \propto e^{-I}$ and it is 
of interest to compare this wavefunction with the 
above ground state wormhole 
wavefunctions. For example, in the superstring 
inspired model, where $\omega =-1$, we find that 
$I= 4a_1 a_2 e^{-\Phi}$. Performing a conformal 
transformation on the four--metric, $\tilde{g}_{\mu\nu} 
= \Theta^2 g_{\mu\nu}$, where $\Theta^2 \equiv e^{-\Phi}$, 
implies that the dilaton field is minimally coupled to gravity in 
the `Einstein--frame', $\tilde{g}_{\mu\nu}$. In terms 
of variables defined in this frame, 
the wavefunction is given by $\Psi \propto \exp(-4 
\tilde{a}_1 \tilde{a}_2)$ and this is 
{\em precisely} the wavefunction 
for the ${\rm R}^3 \times S^1$ quantum 
wormhole that arises in the standard 
Wheeler--DeWitt quantization.  

This is important because 
the ground state of the ${\rm R}^3 \times S^1$ 
quantum wormhole has been 
selected by the 
supersymmetric quantization 
procedure. We emphasize that the 
bosonic component of the supersymmetric wavefunction 
is unique once a solution to the Euclidean 
Hamilton--Jacobi equation has been specified.  
In this sense, therefore, any ambiguities that 
arise in the operator ordering are eliminated. 

Moreover, the interior of a Euclidean 
Schwarzschild black hole has the form of a Kantowski--Sachs 
metric \cite{kl} and it is possible, therefore, that
supersymmetric quantum cosmology may relate a black hole interior 
to a quantum wormhole. 
It would be interesting to consider 
this possibility further. For example,  
such a relationship
would have  
implications for the graceful exit 
problem of the pre--big bang inflationary scenario
\cite{pbb}. 
This problem arises because the classical, dilaton--driven 
inflationary solution becomes singular in a finite 
proper time. At present, no 
generally accepted mechanism has been proposed 
to avoid such a singularity and ensure a
smooth transition to the standard, post--big bang expansion. 
However, an epoch of pre--big bang  
inflation may be formally 
interpreted in the Einstein--frame in terms of gravitational 
collapse \cite{b}. 
If the final state of such a 
collapse were a non--singular supersymmetric
wormhole configuration, such a problem 
could in principle be avoided. It is intriguing that 
whereas 
the pre-- and post--big bang 
branches are related to one another
by the scale factor duality of the classical 
action, the empty fermion 
component of the wavefunction  
is invariant under such a duality transformation.

In principle, 
the supersymmetric quantization 
of other homogeneous, scalar--tensor    
cosmologies can also be considered following 
the method outlined in this paper. 
The Bianchi type II, ${\rm VI}_0$ and ${\rm VII}_0$ 
cosmologies also 
exhibit global symmetries at the 
classical level \cite{clancy} and, in particular, 
the Wheeler--DeWitt equation for the 
Bianchi type II model reduces to Eq. (\ref{unitmass})
after appropriate field redefinitions \cite{H3b}. 
Thus, a similar analysis to that presented in Section IV 
may also be 
performed for this model. 
Similarly, the effective potential 
arising in the Wheeler--DeWitt equation
of the LRS type III model has an 
opposite sign to that given in Eq. (\ref{WDWKS}). 
However, 
the Wheeler--DeWitt equation can be 
transformed into the unit--mass Klein--Gordon 
equation (\ref{unitmass}) after a suitable 
choice  of null variables. 

Finally, there remains the 
open question of the possible relationship 
between the different approaches 
to supersymmetric quantum cosmology. As we 
discussed in the introduction, 
a supersymmetric minisuperspace may be 
obtained directly 
from a 
full four-dimensional $N=1$ supergravity action with the assistance of a 
suitable dimensional reduction for both the bosonic and fermionic 
variables (see, e.g., 
Refs. \cite{paulo,eath,onesusy}). Alternatively, 
a bosonic 
minisuperspace may be extracted from a $(1+0)$--dimensional 
lagrangian and a 
supersymmetric extension established along the lines of 
Refs. 
\cite{witten,graham,Lidsey} or \cite{REV1st,REV2nd}. 
Determining the fundamental similarities and differences   
between these two methods is a 
challenging problem. 
The one attempt to investigate this 
was made in Ref. 
\cite{grahampio}, 
but unfortunately it was  based on an incomplete ansatz for the 
supersymmetric Bianchi type IX model. 
 This particular problem of the ansatz 
was eventually  corrected \cite{paulo,eath,onesusy} 
but 
no further
studies  have been made. 
Such a complex investigation is beyond the objectives and scope of this paper, 
but it would be interesting to consider this topic 
further.

\vspace{0,5cm}
{\small 
\noindent 
{\large\bf Acknowledgements} 

\vspace{0.3cm}

\noindent 
This work  was supported by   
the qrants ESO/INF/1260/98, ESO/PRO/1258/98, 
CRUP-BC N$^{\underline{\rm o}}
$B/73/99 and CERN\-/P/FIS/15190/1999. JEL is supported by the 
Royal Society. PVM is thankful to QMW for hospitality 
where some of this work was completed. 
The authors are   grateful to 
M. Cavagli\`a  and N. Kaloper for helpful 
discussions.}

\end{document}